# THz-to-Optical Conversion in Wireless Communications Using an Ultra-Broadband Plasmonic Modulator

S. Ummethala[1,2,*], T. Harter[1,2], K. Koehnle[1,2], Z. Li[1], S. Muehlbrandt[1,2], Y. Kutuvantavida[1,2], J. Kemal[1], J. Schaefer[4], A. Tessmann[5], S. K. Garlapati[3], A. Bacher[2], L. Hahn[2], M. Walther[5], T. Zwick[4], S. Randel[1], W. Freude[1], C. Koos[1,2,**]

[1]*Institute of Photonics and Quantum Electronics (IPQ), Karlsruhe Institute of Technology (KIT), 76131 Karlsruhe, Germany*
[2]*Institute of Microstructure Technology (IMT), Karlsruhe Institute of Technology (KIT), 76334, Eggenstein-Leopoldshafen, Germany*
[3]*Institute of Nanotechnology (INT), Karlsruhe Institute of Technology (KIT), 76334, Eggenstein-Leopoldshafen, Germany*
[4]*Institute of Radio Frequency Engineering & Electronics (IHE), Karlsruhe Institute of Technology (KIT), 76131 Karlsruhe, Germany*
[5]*Fraunhofer Institute for Applied Solid State Physics (IAF), 79108 Freiburg, Germany*

[*]sandeep.ummethala@kit.edu, [**]christian.koos@kit.edu

**Future wireless communication networks have to handle data rates of tens or even hundreds of Gbit s$^{-1}$ per link, requiring carrier frequencies in the unallocated terahertz (THz) spectrum**[1,2]**. In this context, seamless integration of THz links into existing fiber-optic infrastructures**[3] **is of great importance to complement the inherent portability and flexibility advantages of wireless networks by the reliable and virtually unlimited capacity of optical transmission systems. On the technological level, this requires novel device and signal processing concepts for direct conversion of data streams between the THz and the optical domains. Here, we report on the first demonstration of a THz link that is seamlessly integrated into a fiber-optic network using direct terahertz-to-optical (T/O) conversion at the wireless receiver. We exploit an ultra-broadband silicon-plasmonic modulator having a 3 dB bandwidth in excess of 0.36 THz for T/O conversion of a 50 Gbit s$^{-1}$ data stream that is transmitted on a 0.2885 THz carrier over a 16 m-long wireless link. Optical-to-terahertz (O/T) conversion at the wireless transmitter relies on photomixing in a uni-travelling-carrier photodiode.**

Data traffic in wireless communication networks is currently experiencing explosive growth[4] and will account for more than 60 % of the overall internet traffic by 2021. To meet the associated capacity challenges, wireless communication networks will have to exploit frequency windows of low atmospheric attenuation in the unallocated THz spectrum beyond[1] 0.275 THz. Moreover, future wireless links need to be intimately integrated into fiber-optic infrastructures, for example in terahertz-over fiber (ToF)[3] or fiber-to-the-antenna (FTTA)[5] architectures. This calls for seamless connection of optical fibers to THz transmitter (Tx) and receiver (Rx) front-ends.

At the THz Tx, optoelectronic conversion of data streams from the optical to the THz domain has been demonstrated to offer a variety of advantages[1] over conventional all-electronic approaches. These advantages include wideband tunability of the carrier frequency and the ability to exploit advanced optical circuitry for generation and multiplexing of data streams prior to conversion to the THz domain. Wireless transmission with data rates of 100 Gbit s$^{-1}$ or more have previously been demonstrated[6–8] by direct optical-to-THz (O/T) conversion of a wavelength-division multiplexing (WDM) signal in an ultra-fast uni-travelling-carrier[9] (UTC) photodiode. In contrast to that, direct THz-to-optical (T/O) conversion of data signals at the receiver has not yet been shown, and previous transmission experiments[6–8,10–12] still rely on all-electronic down-conversion[1] of the signals to the baseband using, e.g., sub-harmonic mixers[13] or Schottky diodes[14].

In this paper, we report on the first demonstration of a wireless link that is seamlessly integrated into a photonic network, complementing direct O/T conversion at the THz Tx by direct T/O conversion at the THz Rx. The wireless link operates at a carrier frequency of 0.2885 THz with a maximum line rate of 50 Gbit s$^{-1}$ and bridges a distance of 16 m. The THz signal is generated by O/T conversion in a UTC photodiode. At the receiver, the THz signal is converted to the optical domain by using an ultra-broadband plasmonic-organic hybrid (POH) modulator. The POH modulator features a flat frequency response[15,16] up to 0.36 THz along with small footprint of about 600 µm$^2$, thus lending itself to high-density photonic integration. To the best of our knowledge, this is the first demonstration of direct conversion of a THz wireless data signal to the optical domain without prior down-conversion to the baseband or to an intermediate frequency. We expect that the combination of direct O/T and T/O conversion in ultra-compact devices has the potential to greatly accelerate THz communications and to advance the integration of THz wireless links into fiber-optic infrastructures. The concept relies on distributed THz transceiver (TRx) front-ends that are connected to powerful centralized digital signal processing (DSP) sites through widely deployed fiber-optic network infrastructures that exploit optical carriers to efficiently carry data signals over large distances. The architecture relies on direct O/T and T/O conversion at the TRx front-end, which is key to efficiently interface optical fibers to THz antennae. Direct conversion between analogue optical signals and THz waveforms considerably reduces complexity at the antenna site and improves scalability to a large number of geographically distributed high-performance THz links or cellular networks. Similarly, the concept of moving computationally expensive digital processing of baseband (BB) signals to centralized sites such as large-scale data centers offers unprecedented network scalability, flexible and efficient sharing of crucial resources as well as improved network resilience. Seamless combination of short-reach THz links with long-reach fiber-optic networks may thus represent a key step towards overcoming the capacity bottlenecks in wireless communication infrastructures.

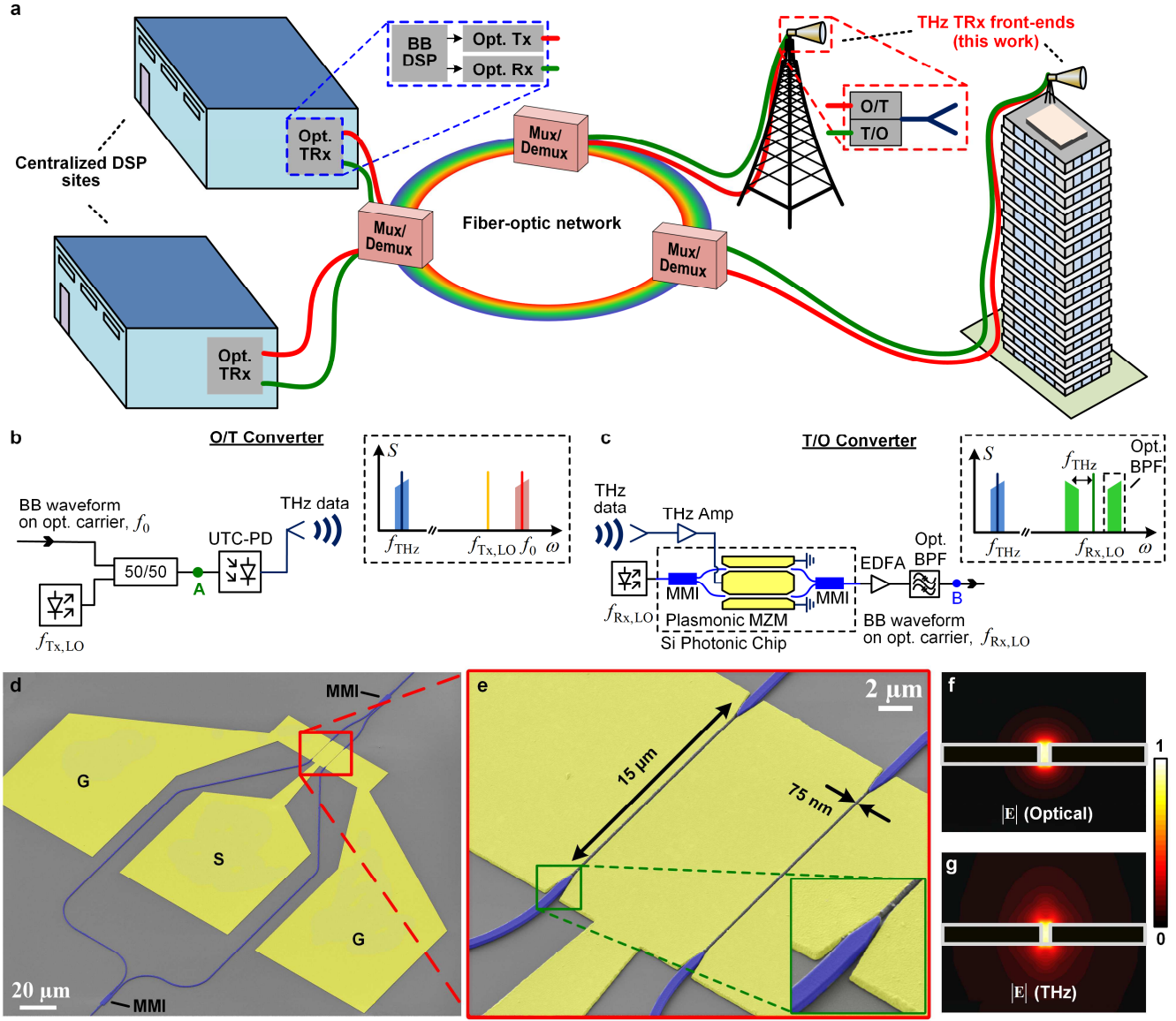

**Fig. 1: Seamless integration of THz wireless links into fiber-optic infrastructures by direct optical-to-THz (O/T) and THz-to-optical (O/T) conversion. a** Vision of a future communication network architecture that combines geographically distributed analogue THz transceiver (TRx) front-ends with powerful centralized digital signal processing (DSP) sites through fiber-optic network infrastructures. Our work focusses on direct T/O and O/T conversion of analogue waveforms at the THz front-end, which allows to efficiently interface optical fibers to THz antennae. This considerably reduces complexity at the antenna site and improves scalability to a large number of THz links or cellular networks. The envisaged concept may thus represent a key step towards overcoming the capacity bottlenecks in wireless communication infrastructures. **b** Concept of direct O/T conversion. The baseband (BB) waveform is first modulated onto an optical carrier at frequency $f_0$ and sent to the O/T converter. The optical signal is then translated to a THz carrier by photomixing with a continuous-wave (CW) local oscillator (LO) laser tone at a frequency $f_{Tx,LO}$ in an ultra-fast uni-travelling-carrier photodiode (UTC-PD). The THz data signal is radiated into free-space by an antenna. The inset shows the optical data spectrum (center frequency $f_0$), the LO tone (frequency $f_{Tx,LO}$), and the THz spectrum (center frequency $f_{THz}$) after O/T conversion. **c** Concept of direct T/O conversion. The THz data signal is received by an antenna, amplified by a THz amplifier, and then fed to a plasmonic Mach-Zehnder modulator (MZM) for modulation onto a CW carrier at optical frequency $f_{Rx,LO}$. After modulation, the optical signal from the MZM contains an upper and a lower modulation sideband. An optical band-pass filter (BPF) is used to suppress the carrier and to select one of the modulation sidebands. The inset depicts an illustration of the spectrum of the THz signal (carrier frequency $f_{THz}$) along with the optical LO tone (frequency $f_{Rx,LO}$) and the optical signal sidebands generated by THz modulation. The BPF is configured to select one of the sidebands (center frequency $f_{Rx,LO} + f_{THz}$). **d** False-colored scanning electron microscope (SEM) image of a plasmonic MZM with ground-signal-ground (GSG) contact pads (yellow) and integrated silicon photonic waveguides (blue) **e** Zoom-in of the plasmonic section of the MZM showing two 15 µm-long phase modulators with photonic-to-plasmonic mode converters (inset). Each section consists of a plasmonic slot waveguide, featuring a 75 nm-wide gap between two extended gold pads (yellow). **f, g** Field profile of the optical quasi-TE mode and the THz field respectively. Both figures indicate the magnitudes of the complex electric field vectors. It can be noticed that the optical and the electronic modes are tightly confined to the slot region, thereby ensuring strong interaction and efficient modulation.

At its heart, the architecture depicted in Fig.1a relies on THz transmitter (Tx) and receiver (Rx) front-ends that allow for direct conversion between optical and THz signals. The underlying concepts are illustrated in Fig.1b and c. After generating the analogue baseband waveform of the data signal at the DSP site, it is modulated onto an optical carrier at frequency $f_0$ by an optical transmitter (Opt. Tx) and then sent to the THz Tx through a fiber-optic network. At the THz Tx, see Fig.1b, the optical signal is then converted to a THz waveform by photomixing with a continuous-wave (CW) local oscillator (LO) tone at frequency $f_{Tx,LO}$ in a UTC photodiode, see Inset of Fig.1b. The THz data signal, centered at the difference frequency $f_{THz} = |f_{Tx,LO} - f_0|$, is then transmitted into free-space by an antenna. At the T/O converter, Fig.1c, the THz data signal is received by another antenna and fed to a THz amplifier. For conversion to an optical carrier, the amplified signal is then coupled to a plasmonic-organic hybrid[17–23] (POH) Mach-Zehnder modulator (MZM), which is fed by an optical carrier at frequency $f_{Rx,LO}$. The MZM generates an upper and a lower modulation sideband. An optical band pass filter (BPF) is used to suppress the carrier and to select one of the sidebands, see inset of Fig.1c. This scheme allows for operation over a wide range of THz frequencies, and does not require any down-conversion to an intermediate frequency before encoding the data onto an optical carrier, and hence considerably reduces the complexity at the THz front-end. After T/O conversion, the analogue signal is sent back through the fiber optic network to an optical receiver (Opt. Rx) in the centralized DSP site.

The scheme illustrated in Fig.1c crucially relies on ultra-broadband electro-optic modulators, offering modulation bandwidths that extend into the THz spectrum. High-speed Mach-Zehnder modulators (MZM) have previously been demonstrated using lithium niobate as the electro-optic medium – either as bulk material[24] or as thin films on silicon[25] or quartz[26] substrates. However, these devices are usually realized in travelling-wave configuration with typical device lengths on the millimeter or centimeter scale, which is not well suited for high-density integration. Moreover, while some lithium niobate devices show measureable sidebands up to modulation frequencies[26] of 0.5 THz, the underlying 3 dB bandwidths[24–26] are so far limited to approximately 0.1 THz. These bandwidth limitations can be overcome by plasmonic-organic hybrid (POH) modulators[17–23] that combine organic electro-optic (EO) materials with ultra-compact plasmonic slot waveguides. The POH concept allows to considerably reduce device footprint and offers a route towards high-density co-integration[22] with advanced silicon photonic circuitry[27,28]. A fabricated POH MZM is shown in the false-colored scanning electron microscope (SEM) image in Fig.1d. Light is coupled to the silicon photonic (SiP) chip via on-chip grating couplers (not shown) and propagates in silicon strip waveguides (blue) as a quasi-transverse electric (quasi-TE) mode. A multimode interference (MMI) coupler splits the light from the input waveguide and launches it into the two arms of an unbalanced MZM. An MMI coupler at the other end of the MZM combines the modulated signals into an output waveguide which is connected to another grating coupler. Each arm of the MZM contains a POH phase modulator section comprising a narrow metallic slot (width $w$ = 75 nm) between the gold electrodes (yellow), see Fig.1e. A pair of tapered silicon waveguides in each arm is used to convert the photonic mode of the silicon strip waveguide to the surface plasmon polariton (SPP) mode in the metallic slot waveguide[29] and vice versa, see inset of Fig.1e. The slots are filled with the organic EO material[30] SEO100. A THz signal applied to the ground-signal-ground (GSG) contacts of the plasmonic MZM leads to a THz electric field in the slots of each of the two arms and thus creates an optical phase shift. Fig.1f and g show that both the optical quasi-TE field and the THz electric field are tightly confined to the plasmonic slot waveguide, leading to a strong overlap and a high modulation efficiency. The MZM is configured to operate in push-pull mode with phase shifts of equal magnitude but opposite signs in each arm. This is accomplished by an appropriate choice of the poling directions[31] of the EO material with respect to the modulating THz field inside the two slot waveguides. Further details on the device fabrication can be found in the Methods.

Due to the strong interaction of the THz field and the optical wave in the plasmonic slot waveguide, POH phase shifters can be very short. This leads to ultra-small parasitic capacitances of the order of a few fF, thereby permitting theoretical RC corner frequencies in excess of 1 THz when connected to a signal source with a 50 Ω internal impedance[18,31]. For our experiments, we have fabricated a 15 µm-long POH MZM with a slot width of 75 nm and characterized its response over an extended frequency range of up to 0.36 THz, limited by the signal sources available in our lab. Fig. 1a shows the basic setup for the bandwidth measurement. The optical CW carrier at a frequency is derived from an external-cavity laser (ECL) and launched into the plasmonic MZM that is driven by a small sinusoidal electric RF or THz signal with varying drive frequency $f_m$. The intensity-modulated optical signal is then amplified by an erbium-doped fiber amplifier (EDFA) and detected by an optical spectrum analyzer to evaluate the phase modulation index $\eta(f_m)$. For each drive frequency, the optical spectrum exhibits a peak at $f_c$ along with two first-order sidebands at $f_c \pm f_m$. Assuming that the MZM is biased at its quadrature (3 dB) point and that both arms of the MZM have the same phase modulation index $\eta$, the sideband-to-carrier power ratio $R_{1,0}$ allows to calculate $\eta$ according to the relation[32] $R_{1,0} \approx \eta^2/4$ for small-signal drive amplitudes $\eta \ll 1$, see Methods and Supplementary Section I for more details. Note that the electric drive power $P_e$ provided by our signal source shows a strong dependency on the modulation frequency $f_m$. Since the phase modulation index $\eta$ is proportional to the electric drive voltage $U_e$, we can eliminate the impact of the frequency-dependent drive power $P_e(f_m)$ by considering the ratio $\eta(f_m)/U_e(f_m)$. Normalizing the fre-quency characteristic of this ratio to its value at a reference modulation frequency of $f_{m,ref} = 2\,\text{GHz}$, we obtain the dimensionless normalized phase modulation index

$$\Gamma(f_m) = 20\log_{10}\left(\frac{\eta(f_m)/U_e(f_m)}{\eta(f_{m,ref})/U_e(f_{m,ref})}\right) \qquad (1)$$

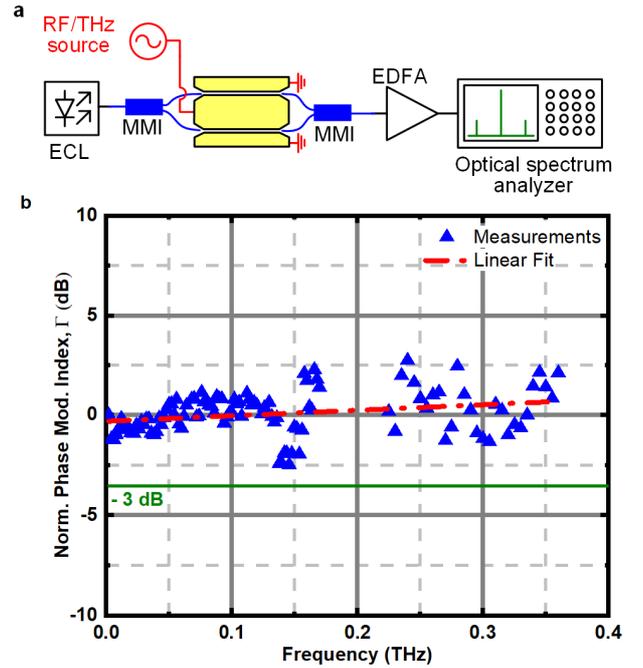

**Fig. 1: Bandwidth measurement of the plasmonic modulator**. **a** Experimental setup for measuring the frequency-dependent phase modulation index $\eta$ of the POH MZM. Continuous-wave (CW) light from an external-cavity laser (ECL) is coupled to the modulator, which is driven by a RF/THz signal. The resulting optical spectrum is recorded by an optical spectrum analyzer and the phase modulation index is extracted from the power ratio of the modulation sidebands and the CW carrier. **b** Frequency dependence of the normalized modulation index $\Gamma$ according to equation (1). The POH MZM features an ultra-broad frequency response up to at least 0.36 THz and shows no indication of bandwidth limitation. The gap around between 0.17 THz and 0.22 THz is due to missing signal source for this frequency range.

Raw data and details on the evaluation can be found in the Supplementary Section I. The results obtained for our 15 µm-long POH MZM are shown in Fig. 2b. The device features a flat frequency response over the entire measurement range with no indication of any frequency-dependent decay. The random variations of the frequency response are attributed to uncertainties in measuring the power of the optical modulation sidebands and in determining the frequency-dependent RF/THz drive power. The experimental results hint to a 3 dB bandwidth that is significantly larger than the highest measured frequency of 0.36 THz. To the best of our knowledge, this represents the fastest EO modulator that has so far been demonstrated. Our findings are well in line with earlier demonstrations of POH modulators, where operation up to frequencies of 0.17 THz has been shown[23], see Supplementary Section I for details. Note that POH devices are not limited to gold as a plasmonic material, but may also be realized by employing CMOS compatible materials[33], which would allow for co-integration with silicon-based electronics.

For the THz wireless transmission experiments, we use a second-generation device with slightly longer MZM (20 µm instead of 15 µm) featuring a slot width of 75 nm and the same electro-optic material (SEO100) as the cladding. For this device, we measure an EO figure-of-merit (FoM) of $n_{EO}^3 r_{33} = 315\,\text{pm/V}$, leading to an estimated EO coefficient of $r_{33} = 64\,\text{pm/V}$. Note that there is vast room to further improve the EO FoM and to hence reduce the drive voltage requirement by using more efficient EO materials. As an example, EO FoM values of $n_{EO}^3 r_{33} = 1990\,\text{pm/V}$ (calculated for $n_{EO} = 1.83$ and $r_{33} = 325\,\text{pm/V}$) have been previously demonstrated in POH devices[34], and even higher values of $n_{EO}^3 r_{33} = 2300 \,\text{pm/V}$ ($r_{33} = 390 \,\text{pm/V}$ for $n_{EO} = 1.81$) have been recently achieved in silicon-organic hybrid (SOH) modulators[35]. Fig. 2a shows photographs of the experimental setups used for the transmission experiment. The THz Tx comprises a high-speed UTC photodiode with a horn antenna to generate the THz signal as well as a plano-convex polytetrafluoroethylene (PTFE) lens that collimates the THz beam. After travelling over a free-space distance of approximately 16 m, the THz signal is coupled to the POH MZM through a second lens, a horn antenna, and a two-stage THz amplifier. A more detailed description of the THz components and a technical sketch of the experimental setup can be found in Supplementary Sections II and III respectively. In the transmission experiment, a quadrature phase-shift keying (QPSK) data stream at a symbol rate of 15 GBd is encoded on an optical carrier at $f_0 = 193.3 \,\text{THz}$ and superimposed with a CW LO tone at $f_{Tx,LO} = 193.3 \,\text{THz}$. Photomixing in the UTC photodiode is used to transfer the optical data stream to the THz carrier frequency of $f_{Tx,THz} = |f_0 - f_{Tx,LO}| = 0.2885 \,\text{THz}$. Fig. 2b shows the optical spectrum at the input of the UTC photodiode, marked as position 'A' in Fig.1b. At the THz Rx, the received signal is boosted by two cascaded THz amplifiers to drive the POH MZM via a ground-signal-ground (GSG) probe with a hollow-core waveguide. The amplifiers are based on millimeter-wave monolithic integrated circuits (MMIC) and feature a total gain of approximately 40 dB for frequencies between 0.270 THz and 0.310 THz, see Supplementary Section II for details. The POH MZM is biased at its quadrature (3 dB) point and hence imposes an intensity modulation on the CW carrier at $f_{Rx,LO} = 194.57 \,\text{THz}$ obtained from a second ECL. The optical data signal is amplified by erbium-doped fiber amplifiers (EDFA) and sent through an optical filter which suppresses the carrier at $f_{Rx,LO}$ along with the upper sideband. The corresponding spectrum taken after the optical filter, i.e., at position 'B' in Fig.1c, is displayed in Fig. 2c. The optical QPSK signal is detected and evaluated using an optical modulation analyzer with an in-built optical LO for coherent intradyne reception. Note that intensity-modulating a THz QPSK signal onto an optical carrier and selective filtering of one of the sidebands generates an optical QPSK signal even though the THz Rx contains only a simple MZM. Note also that, due to the limited tuning range of the optical filter, we chose to isolate the lower sideband of our optical signal rather than the upper one, as sketched in Fig.1c.

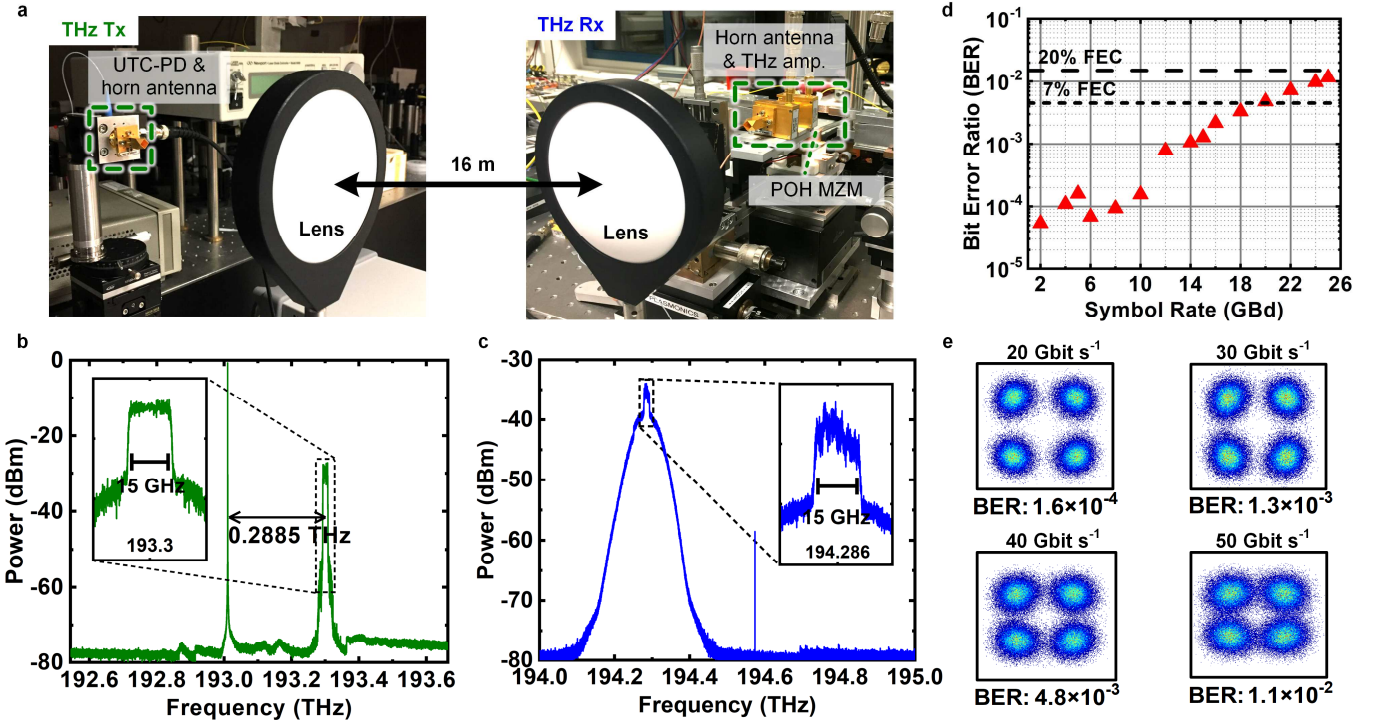

**Fig. 2: Demonstration of THz wireless data transmission using direct O/T and T/O conversion. a** Photograph of the THz transmitter (Tx) and receiver (Rx) setup, separated by a distance of 16 m. The THz Tx comprises a high-speed UTC photodiode (UTC-PD), a horn antenna, and a plano-convex polytetrafluoroethylene (PTFE) lens that collimates the THz beam. After travelling over a free-space distance of approximately 16 m, the signal is coupled to the POH MZM through a second lens, a horn antenna, and a two-stage THz amplifier. A more detailed description and a technical sketch of the data transmission setup can be found in Supplementary Section III and in Supplementary Figure S6. **b** Optical spectrum measured at position 'A' of the THz Tx shown in Fig. 1b. The spectrum consists of a 15 GBd QPSK data signal (inset) at a carrier frequency of 193.3 THz and a CW tone, detuned from the carrier of the data signal by the targeted THz carrier frequency of 0.2885 THz. The THz signal is generated by photomixing in an UTC photodiode. **c** Filtered optical spectrum measured at position 'B' of the THz Rx, see Fig. 1c. Upon modulating the THz signal onto the optical carrier, the upper sideband and the carrier are suppressed by an optical band pass filter such that only the lower sideband and the residual carrier are visible. The lower sideband contains a 15 GBd QPSK signal (inset) centered at 194.286 THz. **d** Measured bit error ratios (BER) of QPSK signals at different symbol rates upon T/O conversion and coherent intradyne detection. The BER stays below the 7 % hard-decision forward error correction (FEC) limit up to a symbol rate of 18 GBd. Higher symbol rates require 20 % soft-decision FEC. **e** Constellation diagrams of the received QPSK signal for data rates of 20 Gbit s⁻¹, 30 Gbit s⁻¹, 40 Gbit s⁻¹, and 50 Gbit s⁻¹.

The QPSK data signal is analyzed offline using standard DSP techniques, and the bit error ratio (BER) is extracted, see Supplementary Section III for details. Fig. 2d shows the BER as a function of the QPSK symbol rate. Up to 18 GBd (line rate 36 Gbit s$^{-1}$), the BER stays below the 7 % hard-decision forward error correction (FEC) limit[36], while we have to resort to a 20 % soft-decision FEC for higher data rates. The BER is mainly limited by the amplified spontaneous emission (ASE) noise of the EDFA. Furthermore, for symbol rates larger than 12 GBd, the BER increases due to a drop in the THz amplifier gain by up to 2.5 dB within the bandwidth of the data signal, leading to signal distortions as well as a reduced drive signal amplitude for the modulator, see Supplementary Section II. Fig. 2e shows the QPSK constellations for line rates of 20 Gbit s$^{-1}$, 30 Gbit s$^{-1}$, 40 Gbit s$^{-1}$ and, 50 Gbit s$^{-1}$ respectively. To the best of our knowledge, this experiment corresponds to the first demonstration of data transmission using direct THz-to-optical conversion.

In summary, we have demonstrated an optical-wireless-optical link with purely optoelectronic frequency conversion both at the THz transmitter and the THz receiver. The link spans a distance of 16 m and is operated at a carrier frequency of 0.2885 THz. Key for this demonstration is a compact ultra-broadband plasmonic modulator offering an unprecedented modulation bandwidth of at least 0.36 THz. With a 20 % soft-decision FEC, we achieve a fiber-to-fiber line rate of 50 Gbit s$^{-1}$. These results demonstrate the prospects of plasmonic modulators as powerful subsystems for THz receiver front-ends. Such devices lend themselves to co-integration with the full portfolio of silicon photonic components that have emerged over recent years[27], possibly complemented by silicon-based nanoelectronic devices[28]. We believe that direct T/O conversion at the Rx can boost terahertz wireless links to data rates of hundreds of gigabit per second.

## Methods

**Fabrication and characterization of plasmonic modulators:** The plasmonic-organic hybrid (POH) modulators used in our experiments are fabricated on standard silicon-on-insulator (SOI) substrates, featuring a 220 nm-thick silicon (Si) device layer and a 2 µm-thick buried oxide (SiO$_2$). Structures are defined by high-resolution electron-beam lithography. A partial Si dry etch step followed by a subsequent full etch of the Si layer is used to form the grating couplers and the 500 nm-wide Si nanowire waveguides. The metallic slots of the plasmonic Mach-Zehnder modulator (MZM) are fabricated via a lift-off process where a 150 nm-thick gold layer is thermally evaporated on a sacrificial layer of poly-methyl methacrylate (PMMA). The metallic slots in the two arms of the MZM are designed to be identical, each featuring a width of $w = 75$ nm, see Fig.1e. The Si nanowire waveguide, depicted in blue in Fig. 1e, ends at the plasmonic section of the MZM with a taper tip angle of 12°, see inset of Fig.1e. These tapers enable efficient conversion between the photonic mode in the silicon waveguide and the surface plasmon polariton (SPP) in the metallic slot with typical conversion losses of only 0.7 dB per transition[17,29]. The silicon waveguides in the two arms of the MZM have a geometrical length difference of 80 µm, see Fig.1d, to enable adjustment of the MZM operating point by tuning the wavelength. The electro-optic cladding consists of the commercially available material SEO100, for which an EO coefficient of $r_{33} = 160$ pm/V has been demonstrated in a thin-film experiment[30]. The material is spin-coated onto the plasmonic MZM and poled by applying a static electric field at an elevated temperature for aligning the randomly oriented EO dipoles[31]. Subsequent cooling preserves the dipole orientation even after removal of the poling field. For a slot waveguide length of $L = 20$ µm, we measured an insertion loss of 16 dB for the plasmonic section. The in-device EO figure-of-merit (FoM) and the EO coefficient are estimated to $n_{EO}^3 r_{33} = 315 \text{ pm/V}$ and $r_{33} = 64 \text{ pm/V}$, respectively, by assuming[30] $n_{EO} = 1.7$. These values are obtained from the measured $U_\pi L$-product of 240 V µm using the relation $U_\pi L = w \lambda_c / (2 n_{EO}^3 r_{33} G)$. In this equation, the parameters $\lambda_c = 1.55$ µm and $U_\pi$ are the operating wavelength, and the measured peak-to-peak modulation voltage that is needed to drive the modulator from constructive to destructive interference for low frequencies. The calculated[31] field interaction factor $G = 0.77$ describes the interaction between the modulating RF field and the optical field. By measuring on-chip silicon nanowire waveguides of different lengths at 1550 nm, we found an insertion loss of 6.5 dB per on-chip grating coupler and a propagation loss of 0.9 dB/mm for the silicon waveguide sections.

**Bandwidth measurement:** The frequency response of the plasmonic modulator is measured by applying a small sinusoidal voltage with variable frequency in the range $2 \text{ GHz} \leq f_m \leq 0.36 \text{ THz}$. The open-circuit transmission-line electrodes of the POH MZM feature a total length of $\sim 150$ µm, see Supplementary Section I for details. For evaluation of the data, we use a global normalization technique that allows to infer the overall shape of the frequency response despite strong variations of the electrical driver power that was available in the various measurement bands. The modulation signal is coupled to the MZM by ground-signal-ground (GSG) probes, which have a wave impedance of $Z = 50 \, \Omega$. The open-circuit peak voltage $U_e$ at the POH modulator is estimated from the electric power $P_e$ which is available at the input of the GSG probe and is measured by replacing the probes by a power meter, see Supplementary Section I for details. To determine the phase modulation index $\eta$, we use a high-resolution optical spectrum analyzer (APEX AP 2050A) to determine the power of the carrier $f_c$ and of the two first-order intensity modulation sidebands at $f_c \pm f_m$. For larger drive amplitudes, a multitude of modulation sidebands would become visible at frequencies $f_c \pm k f_m$ ($k = 1,2,3,...$). Assuming equal phase modulation indices $\eta$ in both arms, the optical power at each frequency is given by[32] $P_o(f_c + k f_m) = P_i J_k^2(\eta)\left(1+(-1)^k \cos \Delta\phi\right)$, where $P_i$ is the power of the light entering the modulator, $J_k(\eta)$ is the $k$-th order Bessel function of the first kind, and $\Delta\phi$ is the phase difference between the two interferometer arms of the MZM that determines the operating point. The sideband-to-carrier power ratio of the first-order sidebands at $f_c \pm f_m$ and the spectral line at $f_c$ is hence given by $R_{1,0} = P_o(f_c \pm f_m)/P_o(f_c) = \alpha J_1^2(\eta)/J_0^2(\eta)$ where $\alpha = (1-\cos(\Delta\phi))/(1+\cos(\Delta\phi))$. For small amplitudes of the modulating signal ($\eta \ll 1$), the Bessel functions can be approximated by $J_0(\eta) = 1$ and $J_1(\eta) = \eta/2$, which leads to $R_{1,0} \approx \alpha\eta^2/4$, see Supplementary Section I. In order to cover the frequency range of interest, we employ various sources: For $2\,\text{GHz} \leq f_m \leq 64\,\text{GHz}$, the source of an Anritsu 37397C vector network analyzer (VNA) is used. A Keysight VNA (PNA-X, N5247) with different frequency extension modules (OML N5262-60003 and V06VNA2-T/R-A) acts as the RF source in the frequency ranges $70\,\text{GHz} \leq f_m \leq 0.11\,\text{THz}$ and $0.11\,\text{THz} \leq f_m \leq 0.17\,\text{THz}$, respectively. The frequency range $0.22\,\text{THz} \leq f_m \leq 0.36\,\text{THz}$ is covered by an electrical signal that is generated by photomixing of two detuned CW laser tones in a uni-travelling-carrier (UTC) photodiode. In the range $0.17\,\text{THz} < f_m < 0.22\,\text{THz}$, no source was available in our laboratory. To eliminate the impact of the frequency-dependent drive signal power, the modulation sidebands are measured together with the available electrical power $P_e$ in front of the GSG probes, using a step size of 2 GHz for frequencies up to 0.17 THz and of 5 GHz in the frequencies from 0.22 THz up to 0.36 THz. A dedicated calibration procedure is used to ensure global comparability of the modulation indices measured in the various frequency bands, see equation (1) as well as Supplementary Section I for details. For frequencies up to 65 GHz, we use an Anritsu ML2438A power meter, whereas the power at higher frequencies is measured by a waveguide-coupled calorimeter (VDI Erickson PM4).

# THz-to-Optical Conversion in Wireless Communications Using an Ultra-Broadband Plasmonic Modulator
# Supplementary Information

S. Ummethala[1,2,*], T. Harter[1,2], K. Koehnle[1,2], Z. Li[1], S. Muehlbrandt[1,2], Y. Kutuvantavida[1,2], J. Kemal[1], J. Schaefer[4], A. Tessmann[5], S. K. Garlapati[3], A. Bacher[2], L. Hahn[2], M. Walther[5], T. Zwick[4], S. Randel[1], W. Freude[1], C. Koos[1,2,**]

[1]Institute of Photonics and Quantum Electronics (IPQ), Karlsruhe Institute of Technology (KIT), Karlsruhe, Germany

[2]Institute of Microstructure Technology (IMT), KIT, Eggenstein-Leopoldshafen, Germany

[3]Institute of Nanotechnology (INT), KIT, Eggenstein-Leopoldshafen, Germany

[4]Institute of Radio Frequency Engineering & Electronics (IHE), KIT, Karlsruhe, Germany

[5]Fraunhofer Institute for Applied Solid State Physics (IAF), Freiburg, Germany

[*]sandeep.ummethala@kit.edu, [**]christian.koos@kit.edu

# Contents

# I. Frequency Response of the Plasmonic-Organic Hybrid (POH) Modulator

We characterize the frequency response of the plasmonic-organic hybrid (POH) Mach-Zehnder modulator (MZM) by measuring the output spectrum of the intensity-modulated light for different modulation frequencies and by determining the power ratio of the first-order modulation sidebands and the optical carrier. To this end, a continuous-wave (CW) optical carrier at optical frequency $f_c$ (angular frequency $\omega_c = 2\pi f_c$) with a power $P_o = 18\,\text{dBm}$ is launched into the POH MZM, which is driven with a small sinusoidal RF/THz signal at a frequency $f_m$, see Fig. 2a of the main paper. The arm lengths of the POH MZM differ by 27 µm, and the wavelength of the optical carrier was chosen to operate the device close to its the quadrature point, where the phase shift in the two arms differs by $\Delta\phi = \pi/2$ in the absence of an electrical modulation signal. The MZM is characterized in the frequency range $2\,\text{GHz} \leq f_m \leq 0.36\,\text{THz}$. Note that this frequency range is only limited by the equipment available in our lab and not by the speed of the POH MZM itself. The intensity-modulated signal is amplified by an erbium-doped fiber amplifier (EDFA) and detected by an optical spectrum analyzer (OSA, APEX AP 2050A, 125 MHz resolution bandwidth) which, for small electric drive signals, reveals a peak at the optical carrier frequency $f_c$ and two first-order modulation sidebands at $f_c \pm f_m$. Assuming that the optical signals at the output of the phase-modulating arms of the MZM feature an identical phase modulation index $\eta$, the sideband-to-carrier power ratio $R_{1,0}$ of the first-order sideband and the optical carrier is given by[1]

$$R_{1,0} = \frac{P_o(f_c \pm f_m)}{P_o(f_c)} = \frac{J_1^2(\eta)(1-\cos\Delta\phi)}{J_0^2(\eta)(1+\cos\Delta\phi)}. \tag{S1}$$

In this relation, $\Delta\phi$ denotes difference of the phase shifts in both arms in the absence of an electrical modulation signal and hence defines the operating point of the device, and $J_0$ and $J_1$ are the 0$^\text{th}$ and the 1$^\text{st}$-order Bessel functions of the first kind. For small electric drive amplitudes $U_e$ of the modulating signal $(\eta \ll 1)$, the Bessel functions can be approximated by $J_0(\eta) = 1$ and $J_1(\eta) = \eta/2$ which leads to

$$R_{1,0} = \alpha\frac{\eta^2}{4}, \quad \alpha = \frac{(1-\cos\Delta\phi)}{(1+\cos\Delta\phi)}. \tag{S2}$$

If the modulator is operated exactly at its quadrature point $\Delta\phi = \pi/2$, then the correction factor $\alpha = 1$. Exact adjustment of the operating point during the experiment, however, turned out to be difficult due to noise and distortions of the cos²-type spectral power transfer function of the unbalanced POH MZM. To reduce the associated uncertainties, we record the power transmission spectrum of the device over an extended spectral range to estimate the exact phase shift $\Delta\phi$ and hence consider the associated correction factor $\alpha$ for the data evaluation. To this end, we fit the measured power transmission spectrum with a model function. We assume that after eliminating the impact of the grating couplers, the power transfer function of the POH MZM is essentially given by $T(\omega) \propto \cos^2(\Delta\phi/2)$ with $\Delta\phi = \beta(\omega)\Delta L$, where $\beta(\omega)$ is the frequency-dependent propagation constant and $\Delta L$ is the path length difference between the two arms of the MZM. For fitting the measured data, we additionally allow for a non-perfect extinction ratio of the MZM, and we adopt a second-order Taylor series expansion of $\beta(\omega)$ about the center frequency $\omega_0$ of the tuning range for taking into account chromatic dispersion. This leads to a model function of the form

$$T(\omega) = A^{(0)} + A^{(1)}\cos^2\left(\frac{1}{2}\left(\beta^{(0)} + \beta^{(1)}(\omega-\omega_0) + \frac{1}{2}\beta^{(2)}(\omega-\omega_0)^2\right)\Delta L\right). \tag{S3}$$

In this relation, $\beta^{(0)}$, $\beta^{(1)}$ and $\beta^{(2)}$ denote the propagation constant, its first and its second-order derivative at the center frequency $\omega_0$. The parameters $A^{(0)}, A^{(1)}, B^{(0)} = \beta^{(0)}\Delta L, B^{(1)} = \beta^{(1)}\Delta L$, and $B^{(2)} = \frac{1}{2}\beta^{(2)}\Delta L$ are determined by a least-square fit of the model function to the measured transmission spectrum, see Fig. S1. Based on these results, the phase shift $\Delta\phi_c = \left(\frac{1}{2}\left(\beta^{(0)} + \beta^{(1)}(\omega_c-\omega_0) + \frac{1}{2}\beta^{(2)}(\omega_c-\omega_0)^2\right)\Delta L\right)$ is determined for each actual operating point, and used to calculate the associated correction factor $\alpha$. Fig. S1 shows the measured power transmission spectrum of the POH MZM as a function of angular frequency $\omega = 2\pi c/\lambda$ along with the associated fit according to the Eq. (S3). Note that the electro-optic frequency response shown in Fig. 2b in the main paper is measured in four different segments, covering the frequency band $2\,\text{GHz} \leq f_\text{m} \leq 64\,\text{GHz}$, $0.07\,\text{THz} \leq f_\text{m} \leq 0.11\,\text{THz}$, $0.11\,\text{THz} \leq f_\text{m} \leq 0.17\,\text{THz}$ and $0.22\,\text{THz} \leq f_\text{m} \leq 0.36\,\text{THz}$, respectively. Each of these measurements are performed on a slightly modified setup, leading to slightly different operating points that are denoted as $P_1$, $P_2$, $P_3 = P_2$, and $P_4$ in Fig. S1.

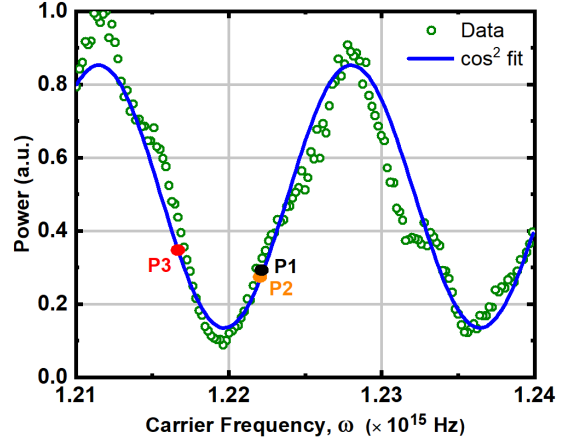

To drive the modulator, we use an external RF/THz source that is connected to the unterminated on-chip coplanar transmission line of the POH MZM using a $50\,\Omega$ ground-signal-ground (GSG) probe. In order to estimate the open-circuit peak voltage $U_\text{e}$ of the electrical signal that is effective at the POH MZM, we first measure the available electrical power $P_\text{e}$ at the input terminals of the GSG probe taking into account losses from various components used in the setup such as cables, connectors, or THz waveguides. For this purpose, we use a calibrated power meter (Anritsu ML2438A) for $2\,\text{GHz} \leq f_\text{m} \leq 64\,\text{GHz}$ and a waveguide-coupled calorimeter (VDI Erickson PM4) for $64\,\text{GHz} \leq f_\text{m} \leq 0.36\,\text{THz}$. We further need to take into account the losses that the THz signal experiences during on-chip propagation along the $150\,\mu\text{m}$-long tapered coplanar transmission line from the probe pads to the POH MZM. To this end, we use a commercially available simulation tool (CST Microwave Studio) to calculate the transmission parameter $S_{21,\text{TL}}$ of the tapered transmission line, see green trace in Fig. S2. Moreover, we need to take into account that the line impedance changes along the transmission line taper from an initial value of $Z_{\text{TL,in}} \approx 50\,\Omega$ at the input to a frequency-dependent impedance $Z_{\text{TL,out}}$ at the output, which is also extracted from the numerical simulation, see blue trace in Fig. S2. Taking into account the insertion loss of the GSG probe $S_{21,\text{Probe}}$ from the datasheet of the manufacturer, the power available at the terminals of the POH modulator can be calculated as $P_{\text{POH}} = P_\text{e} |S_{21,\text{Probe}}|^2 |S_{21,\text{TL}}|^2$. The open-circuit peak drive voltage of the modulating signal at the terminal of the POH MZM is then given by $U_\text{e} = 2\sqrt{2 \times Z_{\text{TL,out}} \times P_\text{POH}}$.

**Fig. S1**: Measured MZM transfer function (green circles) and associated fit (blue line) according to Eq. (S3). $P_1$, $P_2$, $P_3 = P_2$, and $P_4$ denote the operating points used for measuring the frequency response in the bands $2\,\text{GHz} \leq f_\text{m} \leq 64\,\text{GHz}$, $0.07\,\text{THz} \leq f_\text{m} \leq 0.11\,\text{THz}$, $0.11\,\text{THz} \leq f_\text{m} \leq 0.17\,\text{THz}$ and $0.22\,\text{THz} \leq f_\text{m} \leq 0.36\,\text{THz}$ respectively.

To cover the various frequency bands, we use four different GSG probes, which are specified for operation in designated frequency bands: (i) Cascade Microtech Inc. I67-GSG-100 (S/N: HF298) for $2\,\text{GHz} \leq f_\text{m} \leq 64\,\text{GHz}$, (ii) Cascade Microtech Inc. I110-A-GSG-100 (S/N: DV27T) for $0.07\,\text{THz} \leq f_\text{m} \leq 0.11\,\text{THz}$, (iii) Cascade Microtech Inc. I170-T-GSG-100-BT (S/N: HK254) for $0.11\,\text{THz} \leq f_\text{m} \leq 0.17\,\text{THz}$ and (iv) Model 325B from GGB Industries Inc. for $0.22\,\text{THz} \leq f_\text{m} \leq 0.36\,\text{THz}$. The insertion losses ($S_{21,\text{Probe}}$) of the GSG probes are obtained from the respective datasheets. In the frequency range $0.325\,\text{THz} \leq f_\text{m} \leq 0.36\,\text{THz}$, the insertion loss of the THz probe (GGB, Model 325B) is estimated by extrapolating the values from the datasheet, see Fig. S3. Moreover, we need to account for the fact that the drive power $P_\text{e}$ from our signal sources shows a strong dependency with respect to the modulation frequency $f_\text{m}$ and hence results in a strongly frequency-dependent modulation index $\eta(f_\text{m})$. The measured phase modulation index $\eta(f_\text{m})$ and the calculated electrical drive voltage $U_\text{e}(f_\text{m})$ for each frequency band are shown in Fig. S4(a) – (d). Since the phase modulation index $\eta$ is proportional to the electric drive voltage $U_e$, we can eliminate the impact of the frequency-dependent drive power $P_e(f_m)$ by considering the ratio $\eta(f_m)/U_e(f_m)$. Normalizing the frequency characteristic of this ratio to its value at a reference modulation frequency of $f_{m,\text{ref}} = 2\,\text{GHz}$, we obtain the dimensionless normalized phase modulation index according to the Eq. (1) of the main paper,

$$\Gamma(f_m) = 20\log_{10}\left(\frac{\eta(f_m)/U_e(f_m)}{\eta(f_{m,\text{ref}})/U_e(f_{m,\text{ref}})}\right) \tag{S4}$$

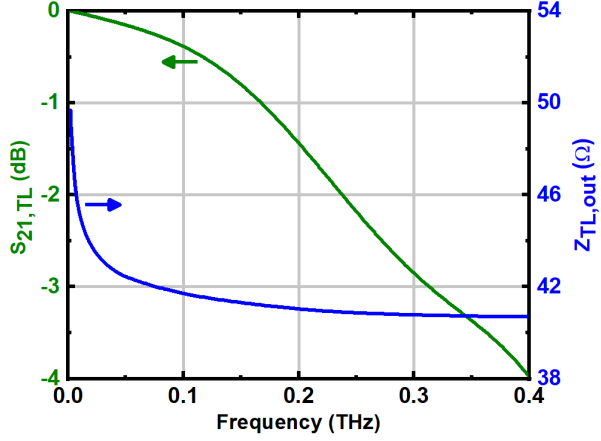

**Fig. S2: Simulated RF/THz characteristics of the on-chip tapered transmission line** The frequency-dependent insertion loss $S_{21,\text{TL}}$ and line impedance $Z_{\text{TL,out}}$ of the tapered coplanar transmission line are obtained using CST Microwave Studio

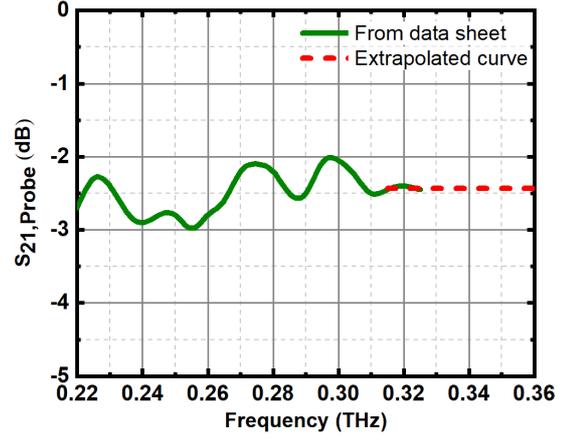

**Fig. S3: Insertion loss of the THz GSG probe.** $S_{21}$ parameter of the THz probe (GGB, Model 325B) as used for calculating the peak electrical drive voltage. The green curve shows the values from the datasheet of the GSG probe and the dashed red curve is an extrapolation in the frequency range between 0.325 THz and 0.36 THz. Note that extrapolation by an essentially flat frequency response is a conservative estimation: assuming a frequency-dependent decay of the $S_{21}$ parameter would lead to a decrease of the drive voltage that is assumed to be effective at the modulator. This would result in an increase of the normalized MZM frequency response in Fig. 2b (in the main paper), that is obtained from dividing the measured phase modulation index by the estimated drive voltage amplitude.

The normalization according to Eq. (S4) allows to infer the global electro-optic response of the POH MZM measured in various frequency bands, leading to the results shown in Fig. 2b in the main paper. From these results it can be observed that the plasmonic MZM has a flat frequency response exceeding 0.36 THz with no sign of bandwidth limitation. Our findings emphasize the fact that the POH modulators are capable of modulating THz frequencies, as predicted in the literature[2,3], and are in line with previous experiments that demonstrated operation of POH modulators[4] at frequencies up to 0.17 THz. Note, however, that in the work presented in ref[4], the frequency response was measured in different frequency bands and the results for each band were then independently normalized to the mean of all data points of the specific band. This technique does not relate the measured modulation index to the actual electric drive amplitude in an extended frequency range and hence makes it difficult to infer the overall frequency roll-off of the modulator transfer function. We overcome this problem by using the global normalization technique presented in the preceding paragraphs.

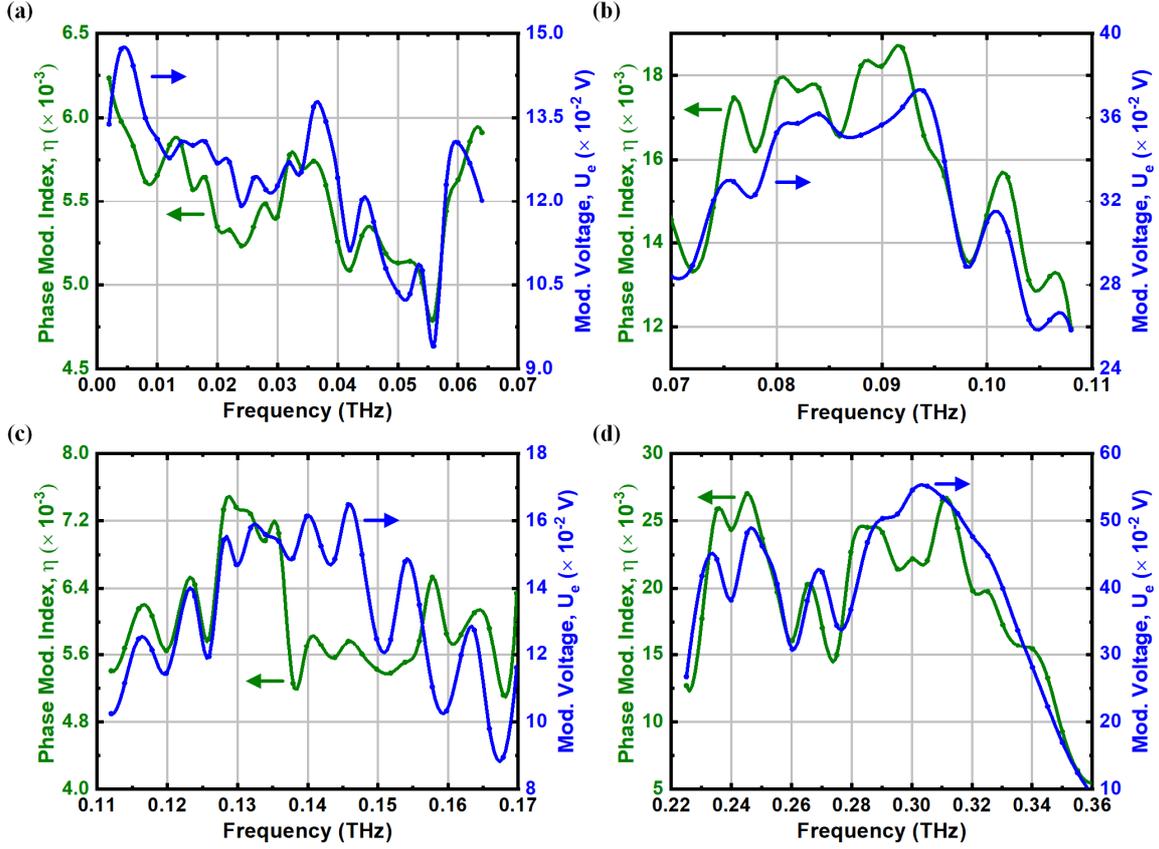

**Fig. S4:** Frequency-dependent phase modulation indices $\eta$ (green traces) of the MZM and associated voltage amplitudes $U_e$ (blue traces) of the electric drive signals for various frequency bands. **(a)** 0.002 THz to 0.07 THz; **(b)** 0.07 THz to 0.11 THz; **(c)** 0.11 THz to 0.17 THz; **(d)** 0.22 THz to 0.36 THz. It can be observed that $\eta$ is essentially proportional to $U_e$, as expected from theory.

## II. Characterization of the Uni-Travelling-Carrier Photodiode and the THz Amplifiers

The frequency response of the uni-travelling carrier photodiode (UTC-PD) employed in the data transmission experiments is characterized using the setup depicted in Fig. S5(a). Light from two detuned external-cavity lasers (ECL, Agilent N7714A) emitting identical polarizations with similar optical powers at frequencies $f_a = 193.0\,\text{THz}$ and $f_b = 193.3\,\text{THz}$ is combined in a polarization maintaining 3 dB coupler. The superimposed signal is amplified to an optical power $P_{\text{opt}}$ and then fed to the UTC photodiode for photomixing to generate a continuous-wave (CW) THz signal at a frequency $f_{\text{THz}} = |f_a - f_b| = 0.3\,\text{THz}$. A polarization controller (PC) is used to adjust the polarization of the superimposed beat signal for maximum THz output signal of the polarization-dependent UTC-PD. The frequency of the CW THz signal is tuned from 0.25 THz to 0.33 THz by varying the optical frequency $f_a$, and the resulting THz output power $P_{\text{THz}}$ is measured with a waveguide-coupled calorimeter (VDI, Erickson PM4). Fig. S5(b) shows the measured THz output power as a function of the generated THz frequency $f_{\text{THz}}$ for an input optical power of $P_{\text{opt}} = 16\,\text{mW}$, which generates a direct photocurrent of $I_{\text{ph}} = 5.6\,\text{mA}$ in the reverse-biased UTC-PD.

Next, we add an H-band medium-power amplifier (MPA, M145AMPH) with a 15 dB gain at 0.3 THz and a saturation output power +4 dBm to the setup to measure the total available THz power after amplification, see Fig. S5(c). The THz amplifier is a millimeter-wave monolithic integrated circuit (MMIC) based on metamorphic high electron mobility transistor (mHEMT) technology[5]. Fig. S5(d) shows the measured THz power as a function of frequency for the combination of UTC-PD and MPA (blue line). Finally, we use the setup depicted in Fig. S5(e) to measure the available THz power after a free-space transmission of 16 m. The UTC-PD transmitter is connected to a horn antenna (26 dBi gain), which radiates the THz signal into free-space. A polytetrafluoroethylene (PTFE) lens is used to collimate the beam that is emitted from the horn antenna. At the receiver, a similar PTFE lens focuses the THz signal to another horn antenna (26 dBi gain). A low-noise H-band amplifier (LNA, M145ALNH) with a gain of 26 dB at 0.3 THz (9 dB noise figure, −30 dBm maximum input power) cascaded with the previously mentioned H-band MPA is used to compensate the free-space loss of the THz signal. The red curve in the Fig. S5(d) shows the THz available power at the receiver after a transmission distance of $d = 16$ m and subsequent amplification by the combination of LNA and MPA. The dips in the red curve are attributed to reflections of the THz carrier due to imperfect waveguide connector flanges between the amplifiers. The operating frequency $f_{\text{Tx,THz}} = 0.2885\,\text{THz}$ of the THz data transmission experiments is marked with a circle and a dashed vertical line. For a perfectly isotropic antenna, the free-space path loss (FSPL) of the wireless link would amount to approximately 105 dB according to the relation $\text{FSPL} = \left(c/4\pi f_{\text{Tx,THz}} d\right)^2$ where $c$ is the vacuum speed of light. From the measurements of the received power, we estimate that the combination of transmitter-receiver lens-and-horn antenna feature a gain of ~ 79 dBi compared to an isotopic antenna, see Section III of the Supplementary Information.

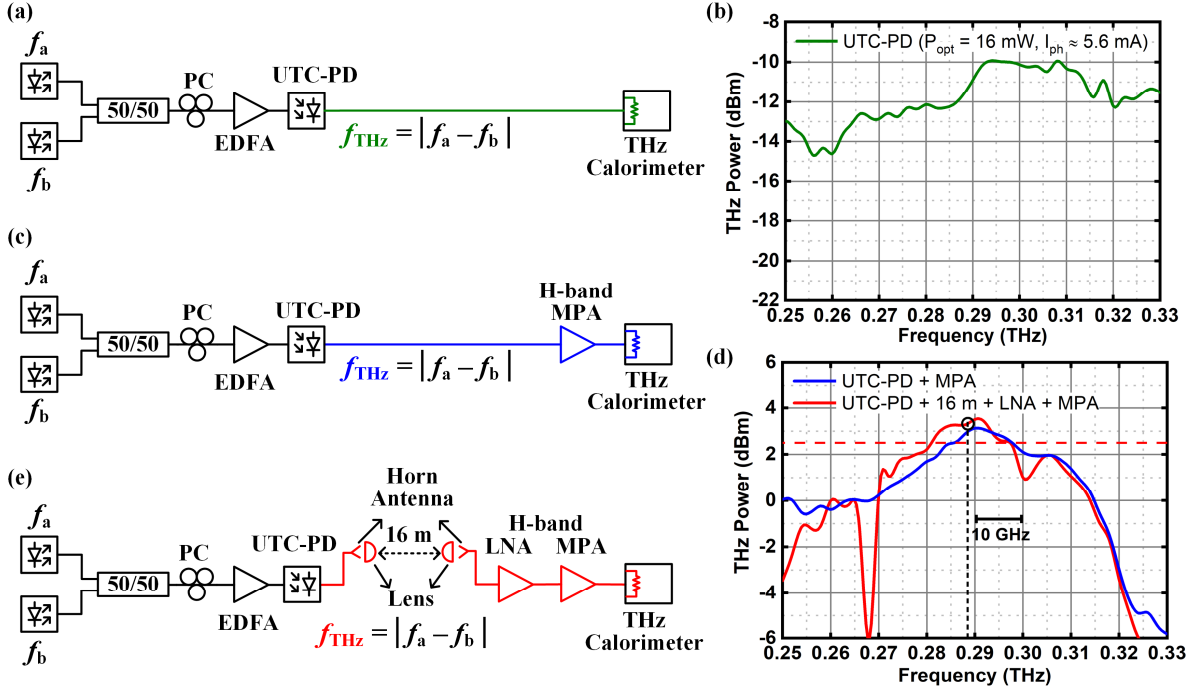

**Fig. S5: Setup for characterizing various THz components used in the data transmission experiment.** For all the measurements, the optical input power to the uni-travelling-carrier photodiode (UTC-PD) is set to $P_{opt} = 16\,\text{mW}$ which corresponds to a direct photocurrent of $I_{ph} = 5.6\,\text{mA}$. **(a)** Setup for measuring the THz output power of the UTC-PD **(b)** Frequency-dependent THz power of UTC-PD as measured by calorimeter **(c)** Setup for measuring the available THz power after amplification by H-band medium-power amplifier (MPA). **(d)** Frequency-dependent THz power for the combination of UTC-PD and MPA (blue curve). **(e)** Setup for measuring the THz power available after a free-space transmission distance of 16 m and a cascaded amplification stage, comprising a low-noise amplifier (LNA) and the MPA. The available THz power as a function of frequency is shown as the red curve in **(d)**. The kinks in the red curve are attributed to imperfect waveguide connector flanges. A horizontal red dashed line indicates THz power of 2.5 dBm, which is needed to achieve a bit error ratio (BER) below the threshold for forward-error correction (FEC) with 7 % coding overhead, see Fig. 3d in the main paper and Fig. S7(b) in Supplementary Section IV. The operating frequency of the THz communication link for the data transmission experiments is marked with a circle and a vertical line.

In the transmission experiment, a minimum THz power of 2.5 dBm is needed at the input of the POH MZM to achieve a BER of $4.5\times10^{-3}$, corresponding to the threshold for forward-error correction (FEC) with 7 % coding overhead. This THz power is indicated by a red dashed line in Fig. S5(d). The usable spectral range with THz output powers above this limit corresponds to a roughly 16 GHz wide band, centered around 0.2885 THz. This explains the increase of the BER with symbol rate as apparent from Fig. 3d of the main paper and eventually limits the data rate that can be transmitted in the current setup.

## III. Experimental Setup Used for Data Transmission Experiments

A detailed sketch of the THz transmitter for optical-to-THz (O/T) conversion is depicted in Fig. S6(a). A software-driven digital-to-analog converter (DAC, Keysight M9505A, sampling frequency 90 GSa s$^{-1}$) is used to generate drive signals for QPSK modulation using a pseudo-random binary sequence (length $2^{11} - 1$). The signals are shaped to Nyquist pulses featuring a raised-cosine spectrum and a roll-off factor of $\beta = 0.1$. The resulting analog signal is fed to an optical IQ-modulator that is fed by continuous-wave (CW) carrier generated by an external-cavity laser (ECL) at an optical frequency $f_0 = 193.3 \text{ THz}$. The IQ-modulator is biased at its zero-transmission point in order to suppress the optical carrier at frequency $f_0$. The intensity-modulated optical signal is amplified using an erbium doped fiber amplifier (EDFA). Out-of-band amplified spontaneous emission noise from the EDFA is blocked by an optical band-pass filter (0.6 nm passband width) before superimposing the modulated optical carrier with a CW local-oscillator tone at a frequency $f_{\text{Tx,LO}}$ derived from another ECL. The power of the CW tone at $f_{\text{Tx,LO}}$ is adjusted to the same level as the average power of the modulated signal. The combined signal is further amplified by a second EDFA, polarization-adjusted, and then fed into a variable optical attenuator (VOA). The CW tone and the optical data signal are mixed in a UTC-PD, resulting in a THz QPSK data stream at a carrier frequency $f_{\text{Tx,THz}} = |f_0 - f_{\text{Tx,LO}}| = 0.2885 \text{ THz}$. The polarization of the superimposed signal entering the polarization-sensitive UTC-PD is adjusted to maximize the THz output power. This THz transmitter implementation provides the flexibility of tuning the THz carrier frequency $f_{\text{Tx,THz}}$ over a broad frequency range, only limited by the bandwidth of the UTC-PD, by adjusting the frequency $f_{\text{Tx,LO}}$ of the unmodulated LO tone. The THz output power provided by the UTC-PD typically amounts to −11 dBm. A conical horn antenna with a gain of 26 dBi and an aperture diameter of 5.6 mm is attached directly to the UTC-PD to radiate the THz signal into free-space. The radiated signal is collimated by a plano-convex PTFE lens (diameter – 101.6 mm and focal length – 200 mm) and is sent over a wireless transmission distance of 16 m.

Fig. S6(b) shows the THz receiver setup used for THz-to-optical (T/O) conversion. Light emitted by an ECL at a frequency $f_{\text{Rx,LO}}$ is amplified to a power of 18 dBm by an EDFA and then launched into the POH MZM using an on-chip grating coupler on the silicon (Si) photonic chip. The on-chip insertion loss of the MZM with 20 µm-long POH phase shifters amounts to 16 dB. The grating couplers and the on-chip waveguides contribute to an additional loss of 13 dB, with vast potential for further reduction by using optimized passive silicon photonic devices. For wireless reception, the THz QPSK data stream is focused to a horn antenna (26 dBi gain) by a PTFE plano-convex lens similar to the one used in the THz transmitter setup. A cascaded combination of H-band LNA and MPA boosts the power of the THz signal and drives the POH MZM through a THz GSG probe (GGB picoprobe Model 325B). The intensity-modulated signal out of the plasmonic modulator is coupled out of the chip using another grating coupler and then amplified by an EDFA. One of the modulated side bands of the amplified signal at frequencies $(f_{\text{Rx,LO}} \pm f_{\text{Tx,THz}})$ is filtered with a 0.6 nm band-pass filter and sent through a second amplification stage, where the optical signal is further amplified to a power of 1 dBm. The resulting signal is again filtered with a 1 nm band-pass filter to additionally suppress the optical carrier and one of the modulation sidebands before being detected by a coherent optical receiver (Agilent optical modulation analyzer N4391A, OMA). The received electrical signal is recorded by a real-time oscilloscope (32 GHz analog bandwidth) and evaluated off-line using digital signal processing (DSP). After resampling and clock-recovery, the recorded data is equalized using a constant-modulus algorithm. The phase of the carrier is estimated based on a Viterbi-Viterbi algorithm, and the bit error ratio (BER) is computed as a final step.

Fig. S6(c) shows a photograph of the THz communication link with the transmitter and receiver separated by 16 m. The THz beam path contains a horn antenna and a plano-convex PTFE lens on each side. Fig. S6(d) shows a photograph of the THz receiver, comprising a horn antenna and a combination of H-band LPA and MPA which drive the POH MZM through a THz ground-signal-ground (GSG) probe. From the free-space path loss (FSPL) of approximately 105 dB, see Section II of the Supplementary Information, we estimate that the directivity gain of each antenna-lens combination is approximately 40 dBi. We later realized that the insertion loss of the THz beam path could be further reduced by more careful alignment of the components, leaving room for further increasing the performance of the transmission link.

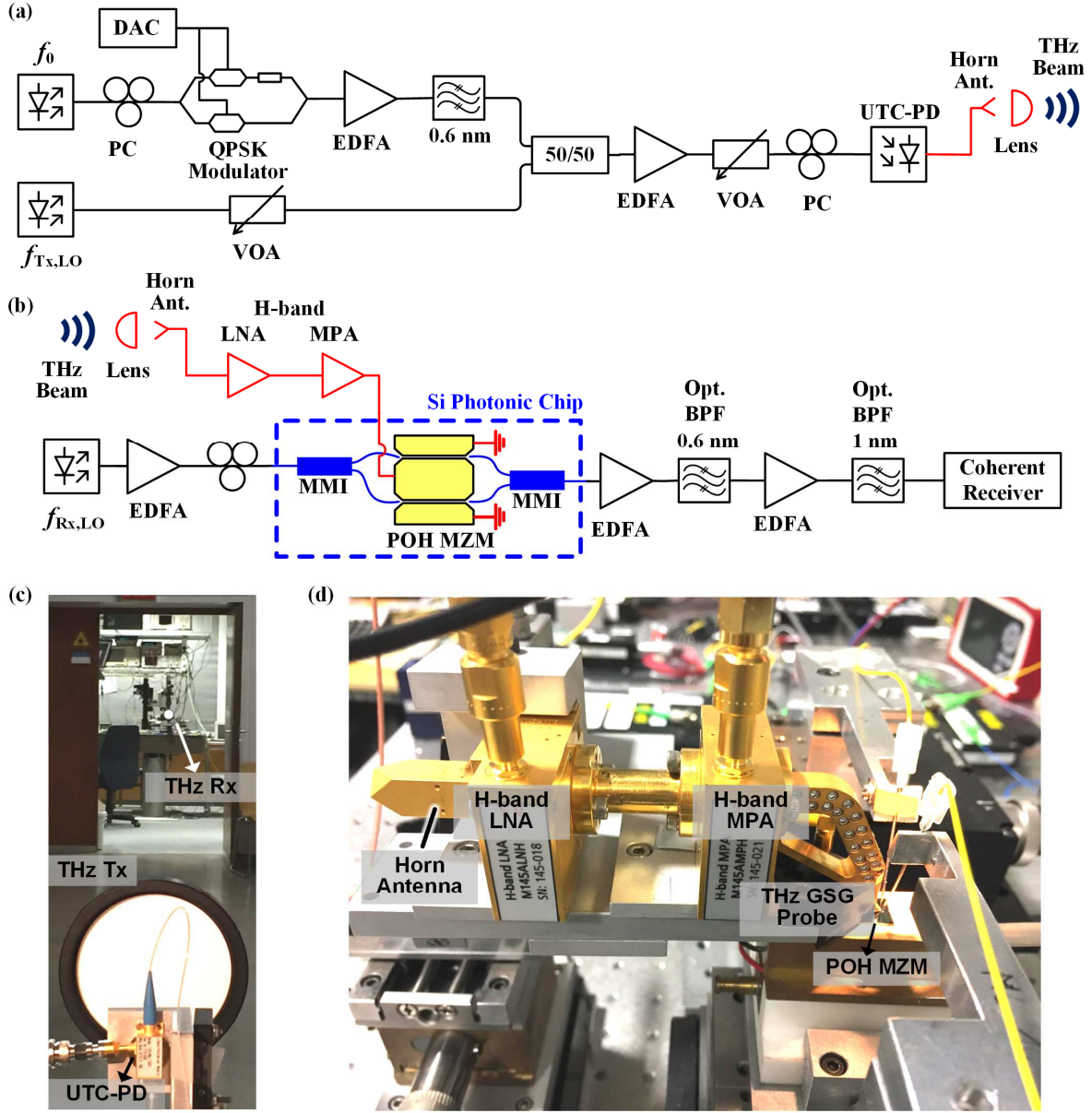

**Fig. S6: Detailed setup of the THz transmitter and receiver used for data transmission experiments.** (**a**) Setup of the wireless terahertz transmitter that generates a QPSK data signal on a carrier frequency of $f_{\mathrm{Tx,THz}} = 0.2885\,\mathrm{THz}$ by exploiting photomixing in a uni-travelling-carrier photodiode (UTC-PD) for optical-to-THz (O/T) conversion. The THz signal is radiated into free-space by a horn antenna. A subsequent polytetrafluoroethylene (PTFE) lens is used to collimate the THz beam. (**b**) Setup for terahertz-to-optical (T/O) conversion at the THz receiver using a plasmonic Mach-Zehnder modulator (MZM). The incoming signal is focused into a horn antenna by a second PTFE lens. It is then fed to a cascade of THz (H-band) amplifiers, which drive the POH MZM, thereby modulating the signal onto an optical carrier. A cascade of EDFA and optical band-pass filters (BPF) is used to amplify the intensity-modulated optical signal and to isolate one of the sidebands. Note that, due to the limited tuning range of the optical filter used in this experiment, we chose to isolate the lower sideband of our optical signal rather than the upper one, as sketched in Fig.1c of the main paper. (**c**) Photograph of the data transmission setup showing the THz Tx and the Rx. Tx and Rx are built up in different laboratories, separated by 16 m of free-space transmission distance across the hallway. (**d**) Photograph of the THz photonic Rx. The wireless signal is focused by a PTFE lens (not shown), then coupled into the horn antenna and amplified by the dual-stage H-band amplifier, which drive the POH MZM through the ground-signal-ground (GSG) probe. DAC – digital-to-analog converter, PC – polarization controller, MMI – multimode interference coupler, EDFA – erbium doped fiber amplifier, VOA – variable optical attenuator, LNA – low noise amplifier, MPA – medium power amplifier.

## IV. Back-to-Back Data Transmission Measurements

To test the transmission setup prior to the wireless transmission experiment, we performed a back-to-back measurement. To this end, we directly connect the THz transmitter (UTC PD) to the POH MZM using only the H-band medium-power amplifier (MPA), see Fig. S7(a). For B2B measurements, $f_{\text{Tx,LO}}$ is set to 193.006 THz resulting in a THz transmitter frequency $f_{\text{Tx,THz}} = |f_0 - f_{\text{Tx,LO}}| = 0.294$ THz . Fig. S7(b) shows the evaluated bit error ratios (BER) for different symbol rates of the QPSK signal after offline DSP of the recorded data. A BER of $5.4 \times 10^{-5}$ is recorded for a QPSK symbol rate of 2 GBd (line rate 4 Gbit s$^{-1}$). The BER worsens and reaches the threshold forward error correction (FEC) schemes with 7 % overhead[6] for a symbol rate of 8 GBd (line rate of 16 Gbit s$^{-1}$). For a 15 GBd QPSK signal (line rate of 30 Gbit s$^{-1}$), the signal quality further deteriorates slightly below the 20 % FEC limit. The BER is limited by the amplified spontaneous emission (ASE) noise of the EDFA as well as by the drop in the transmitter power for signals with larger bandwidth, see Supplementary Section II and blue curve in Fig. S5(d).

Note that the setup for the back-to-back test is different from the one used in the actual transmission experiment, in which the received signal was amplified by a cascade of the H-band MPA and the H-band low-noise amplifier (LNA). Using the amplifier cascade was impossible for the back-to-back test, since this would have led to excessive input power at the MPA. As a consequence, it is impossible to directly compare the results. In fact, for a given symbol rate, the BER obtained from the back-to-back test is even higher than that of the transmission experiment. The reason is apparent from Fig. S5(d): At a carrier frequency of 0.2885 THz indicated by the dashed vertical line, the THz output power available after the cascade of the UTC-PD, the free-space link, and the combined LNA/MPA assembly is larger than the THz power available after the MPA in the back-to-back test. Since noise at the optical receiver is dominated by amplified spontaneous emission (ASE) of the EDFA after the POH MZM, the higher available THz drive power of the MZM directly translates into a lower BER for the transmission experiment as compared to the back-to-back test.

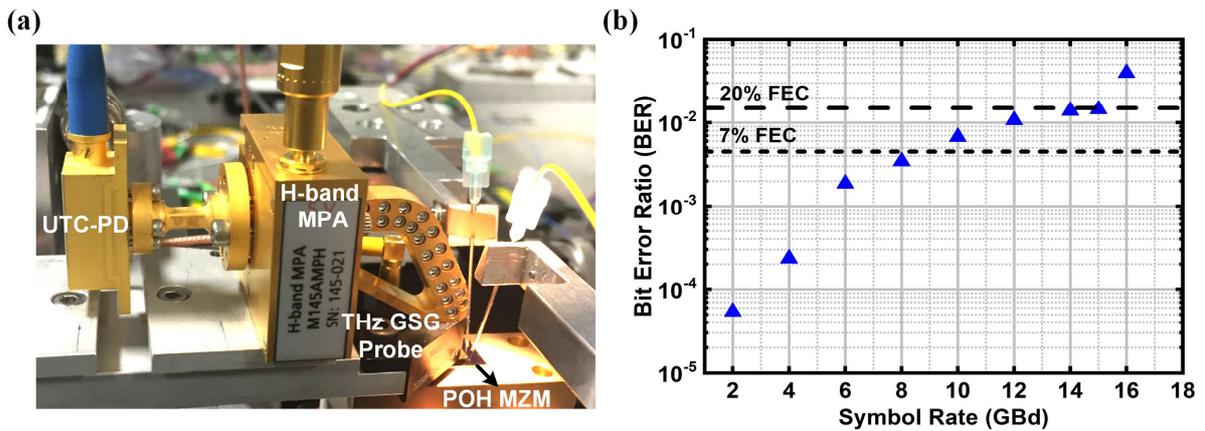

**Fig. S7: Back-to-back data transmission experiment. (a)** Photograph of the measurement setup used for the back-to-back data transmission. The THz data signal from the UTC-PD is amplified using an H-band medium-power amplifier (MPA) and directly coupled to the plasmonic MZM via a GSG probe. **(b)** Measured bit error ratios (BER) of the back-to-back data experiment for different symbol rates of THz QPSK signal. Note that the setup for the back-to-back test is different from the one used in the actual transmission experiment, in which the received signal is amplified by a cascade of the H-band MPA and the H-band low-noise amplifier (LNA). It is hence impossible to directly compare the results – for a given symbol rate, the BER obtained from the back-to-back test is even higher than that of the wireless transmission experiment.